\documentclass[aps,pra,twocolumn,showpacs,groupedaddress,letterpaper]{revtex4-1}

\usepackage{graphicx}
\usepackage{subfigure}
\usepackage{color}
\usepackage{physymb}
\usepackage{enumitem}
\usepackage{mdframed}
\usepackage{framed}
\usepackage{epstopdf}
\newcommand{\codetable}{
\begin{table}[h]
\caption{Code scheme used for the third stage of the interview coding. The \emph{Environment} codes were \emph{a priori} where as the \emph{Experiences} and \emph{Evaluations} codes were the result of the iterative coding described in section \ref{sec:coding}.}
\begin{ruledtabular}
\begin{tabular}{p{2.5cm} p{6.2cm}}
Code & Definition \\ \hline
\emph{Environment} & \\
{\bf Undergrad Course} & Interviewee discussed events that occurred during a non-introductory undergraduate physics lab course. \\
\rule{0pt}{4ex}{\bf Undergrad Other} & Interviewee discussed events that occurred in a research experience or internship during or shortly after undergraduate (prior to graduate school). \\
\rule{0pt}{4ex}{\bf Grad Research} & Interviewee discussed events that occurred during their graduate research experience (Masters or Ph.D.).\\ \hline
\emph{Experiences} &  \\
{\bf Instruction} & Received some form of written or verbal instruction/training in how to keep/use a lab notebook. (\emph{e.g.}, written guidelines, lectures on notebook use, or lab references)  \\
\rule{0pt}{4ex}{\bf Practice} & Practiced using a lab notebook (writing in and/or referencing) for records germane to lab activities. (\emph{e.g.}, notebook use in lab course, during URE, or for graduate research projects) \\
\rule{0pt}{4ex}{\bf Feedback} & Received some form of written or verbal feedback on their use of a lab notebook. (\emph{e.g.}, grading of notebook in course, PI/supervisor provides suggestions about content or organization of notebook) \\ \hline
\emph{Evaluations} & \\
{\bf Adequate} & Experience described as being adequate/sufficient for learning how to maintain a notebook in an authentic research setting.\\
\rule{0pt}{4ex}{\bf Beneficial} & Experience described as being beneficial in learning how to maintain a notebook in an authentic research setting, but not adequate enough for developing their current practice.\\
\rule{0pt}{4ex}{\bf Not Beneficial} & Experience described as not being beneficial in learning how to maintain a notebook in an authentic research setting.\\
\end{tabular}
\end{ruledtabular}
\label{tab:codes}
\end{table}
}

\newcommand{\resultstable}{
\begin{table*}[t]
\caption{Results of the third stage of coding as described in Section \ref{sec:methods}. A (= adequate), B (= beneficial), N (= not beneficial), and `-' (= not present) correspond to the codes in Table \ref{tab:codes}. $^\dagger$Occurred during separate masters program. $^\star$Did not involve interviewee's Ph.D. advisor.}
\begin{ruledtabular}
\begin{tabular}{l | c c c | c c c | c c c}
 & \multicolumn{3}{c |}{\underline{Undergrad Courses}} & \multicolumn{3}{c |}{\underline{Undergrad Other}} & \multicolumn{3}{c }{\underline{Grad Research}} \\
Interviewee & Instruction & Practice & Feedback & Instruction & Practice & Feedback & Instruction & Practice & Feedback \\ \hline
I01 & A & A & A & - & - & - & \:\:A$^\dagger$ & A & \:\:A$^\dagger$ \\
I02 & - & A & - & - & B & - & B & A & B \\
I03 & N & B & N & - & A & - & \:\:N$^\star$ & A & - \\
I04 & - & B & - & - & B & - & - & B & - \\
I05 & - & B & - & - & A & - & - & A & - \\
I06 & N & N & N & - & - & - & - & A & - \\
I07 & N & N & N & - & - & - & \:\:A$^{\dagger\star}$ & A & - \\
I08 & - & N & - & - & B & - & - & B & - \\
I09 & - & N & - & - & B & - & \:\:A$^\star$ & A & - \\
I10 & - & - & - & - & - & - & \:\:B$^\star$ & A & - \\
I11 & - & - & - & - & A & - & - & A & - \\
I12 & - & - & - & B & B & - & \:\:B$^\star$ & B & - \\
I13 & - & - & - & B & B & - & - & A & - \\
\end{tabular}
\end{ruledtabular}
\label{tab:coderes}
\end{table*}
}

\begin{document}

\title{Lab notebooks as scientific communication: investigating development from undergraduate courses to graduate research}

\author{Jacob T. Stanley}
\affiliation{Department of Physics, University of Colorado, Boulder, CO 80309, USA}

\author{H. J. Lewandowski}
\affiliation{Department of Physics, University of Colorado, Boulder, CO 80309, USA}
\affiliation{JILA, National Institute of Standards and Technology and University of Colorado, Boulder, CO, 80309, USA}

\begin{abstract}
In experimental physics, lab notebooks play an essential role in the research process. For all of the ubiquity of lab notebooks, little formal attention has been paid to addressing what is considered `best practice' for scientific documentation and how researchers come to learn these practices in experimental physics. Using interviews with practicing researchers, namely physics graduate students, we explore the different experiences researchers had in learning how to effectively use a notebook for scientific documentation. We find that very few of those interviewed thought that their undergraduate lab classes successfully taught them the benefit of maintaining a lab notebook. Most described training in lab notebook use as either ineffective or outright missing from their undergraduate lab course experience. Furthermore, a large majority of those interviewed explained that they did not receive any formal training in maintaining a lab notebook during their graduate school experience and received little to no feedback from their advisors on these records. Many of the interviewees describe learning the purpose of, and how to maintain, these kinds of lab records only after having a period of trial and error, having already started doing research in their graduate program. Despite the central role of scientific documentation in the research enterprise, these physics graduate students did not gain skills in documentation through formal instruction, but rather through informal hands-on practice.
\end{abstract}

\maketitle

%%%%%%%%%%%%%%%%%%%%%%%%%%%%%%%%%%%%%%
\section{\label{sec:introduction}Introduction}

Scientific communication is commonly defined as the communication of scientific results to the community, typically through scientific presentations and publications in scientific journals. These types of communication can be thought of as the final stages of the scientific process. The physics education research community has devoted little attention to the role these skills play in physics students' education\cite{Franklin2014,Rodriguez2012}. However, another facet of scientific communication, which has received essentially no research attention, is the process of scientific documentation---the communication of data, interpretations, ideas, and results that occur incrementally throughout the process of experimental physics. The need for research in this area has been expressed in a recent report by the American Association of Physics Teachers (AAPT)\cite{AmericanAssociationofPhysicsTeachers2014}, as well as by the President's Council of Advisors on Science and Technology\cite{PCAST2012}, and an earlier report by the National Science Foundation (NSF)\cite{NSF1996}. These reports have emphasized the importance of undergraduate students' engagement in activities similar to those of professional scientists as a priority of undergraduate education. The use of laboratory notebooks for documenting scientific records has been outlined as one such skill.

Scientific documentation is a form of communication that occurs between researchers and their colleagues or advisors, and is something researchers take part in on a daily basis. These scientific records, found in notebooks in experimental physics research labs at universities around the world, are the intellectual foundation of publications in physics journals and talks given at conferences. Given that physics graduate students are some of the primary contributors to these records, and that developing these skills in undergraduate courses has been articulated as a national learning goal, we think it most prudent to focus our attention on exploring how these researchers develop their documentation skills.

Though this type of communication certainly occurs in other academic disciplines and in the broader landscape of industry, we expect these environments might have markedly different standards and norms for how scientific documentation is taught and practiced. Therefore, we chose to narrow our focus to studying the practices of physics graduate students, involved in experimental research. 

In physics, there have been some pedagogical efforts by instructors to improve students' practice of scientific documentation\cite{Dayton1962a,Miller1965a,Atkins2013,Salter2013} as well as some historic case studies of the notebooks of high profile scientists\cite{Johnson1960,McKnight1967}. However, we are unaware of studies that attempt to understand the current state of scientific documentation in experimental physics, by studying the educational trajectories of current graduate student researchers, as we do here. 

The kind of documentation we are focused on is defined by the following characteristics: (1) the recorded information is the result of in the moment thinking, during lab activities; (2) the records are written or referenced at frequent intervals by some or all of the researchers involved in the experiment; (3) the records track the evolution of the experiment and act as the definitive account of what transpired; and (4) the records are utilized to produce more summative forms of communication, such as publications, presentations, and reports. These features define this documentation as a formative form of communication. In research, this documentation may be found in various formats such as bound paper notebooks, electronic notebook programs (e.g., Evernote), or on-line cloud-hosted documents (e.g., Google Docs or a blog). 

Traditionally, one of the earliest points that students may be exposed to this process of scientific documentation is in their undergraduate lab courses---they record experimental information during weekly lab activities. Ostensibly, this helps them make sense of the course content and produce lab reports. Therefore, these courses could ideally serve as early training for the development of these communication skills. However, lab courses may have a range of different learning goals, and development of scientific documentation may or may not be one of them. So, it is unclear whether or not in this context students effectively learn these skills, or if there are other experiences in undergraduate and graduate school that contribute significantly to their development. Given the lack of existing research on this topic, our work takes an exploratory approach to understand the different pathways through which physics students, who end up in graduate school, develop these documentation skills. This will provide a view of the current state of these skills, and may help to inform future educational efforts.

This study addresses several questions about physics graduate students' practice and development of scientific documentation skills with lab notebooks. Specifically: (1) What experiences did they have with lab notebooks in their non-introductory undergraduate lab courses and how do they evaluate the quality of these experiences? (2) Did they have any other experiences with scientific documentation during their undergraduate education? (3) What kinds of guidance, practice, or feedback did they receive from their advisors or colleagues in their graduate program? (4) How do they evaluate the current state of the records they document?  Essentially, we are interested in their educational experiences with lab notebooks and how this has led to the graduate students' current practice. Here, we outline the the development of these skills, and highlight some of the more common experiences at different stages of education. In order to do this, we conduct and analyze interviews of physics graduate student researchers.

%%%%%%%%%%%%%%%%%%%%%%%%%%%%%%%%%%%%%%%%%
\section{\label{sec:methods}Methods}

In this section, we provide a description of the the methodological approach we took with our study. Additionally, we provide details of the data collection, including the selection criteria for the interviewees and the structure of the interview protocol. Finally, we outline how the data were analyzed.

\subsection{Methodological Approach}

Our aim was to probe and understand physics graduate students' educational experience involving the use of lab notebooks (both in undergrad and grad programs) and learn how these experiences led to their current practice of scientific documentation. Because of the limited amount of research done into how graduate students learn to perform scientific documentation, we took an exploratory approach to this study. We conducted semi-structured interviews of physics graduate students, currently participating in experimental research. The interview protocol primed the interviewees to reflect on, describe, and evaluate various incidents that potentially involved lab notebook use: undergraduate lab courses, undergraduate research experiences, internships, graduate research, etc.

Through an iterative coding process we identified broad experience categories, in order to quantify the interviewees' use of lab notebooks during the stages of their education. Furthermore, we categorized their evaluations of these experiences in a limited number of ways, so we could identify whether or not the interviewee found the experience to be helpful in improving their documentation skills. This provided us with a overview of the role lab notebooks played in the education of our participants.

During the data analysis, several specific themes emerged that were common to many of the interviewees. We then sought to capture and quantify these by tallying the number of interviewees whose described experiences could be categorized into each of the themes. We bolstered these quantitative results with qualitative presentation of quotations that detail how the interviewees described the identified themes. In doing so, we provide a richer depiction of our simpler quantitative findings.

It should be noted, we acknowledge that we did not attempted to distinguish between individuals' conceptualizations of past events and the actual experiences of the events themselves. In other words, we have accepted the statements made by participants, without questioning their credibility, while acknowledging that retrospective bias is present in these recollections. This is a common limitation for studies that rely on recollection of past events, and should be taken into consideration by the reader. Despite this limitation, we believe our results provide a fruitful starting point to understand how physics researchers develop their scientific documentation skills. 

Details of the interviewee selection, interview protocol, and the coding/analysis are described in Section \ref{sec:intselection} through Section \ref{sec:coding}.

%%%%%%%%%%%%%
\subsection{\label{sec:intselection}Interviewee Selection}

The interviewees for this study included 13 physics graduate students (six women, seven men), at the University of Colorado Boulder. The primary criterion for inclusion was that they were actively involved in research in one of the ``table-top'' experimental physics subfields (i.e., those that involve daily, hands-on experimental activities with in-house equipment). The breakdown of number of interviewees and their disciplines were seven in Atomic, Molecular, and Optical physics (AMO); three in Condensed Matter physics (CM); two in Biophysics; and one in Plasma physics. We chose to focus on these sub-fields because the daily activities and hands-on nature of the research are most in-line with the activities that students would be involved in during a non-introductory lab course, and thus the use of notebooks for documentation were likely to be similar. The pool of interviewees represented nine different research groups, with no more than two interviewees coming from any one group. Furthermore, all interviewees had to have spent at least six months doing research in their current group. The interviewees represented a range of years---the most junior being in their second year and the most senior being in their sixth. Finally, all interviewees completed an undergraduate physics degree in which they had at least one non-introductory physics lab course during their undergraduate physics education. Four interviewees went to small (fewer than 3,000 students) private liberal arts colleges for their undergraduate education; two went to small to mid sized (fewer than 10,000 students) private research universities; three went to large (greater than 10,000 students) private research universities; and four went to large (greater than 10,000 students) public research universities.

The interviewees were paid volunteers who responded to an email request for research participants. The email stated that we were interested in getting feedback from physics graduate students about their use of notebooks in their research for the purposes of a physics education research study. Furthermore, we specified that the interview was not intended as a personal assessment of their use of lab notebooks, but rather a way to collect information about the range of lab notebook use in an authentic research setting. We emailed a total of 30 individuals. Of those, 16 agreed to be interviewed (three of which were ultimately removed from the data pool, as described in Section \ref{sec:pre-code}).

The generalizability of the results of this study may or may not extend to other large public doctoral universities due to the fact that all participants are graduate students at University of Colorado Boulder. Aspects, such as the size and selectivity of the physics program, type of research being done in the department, and the racial or ethnic demographics of the institution and student body, will factor in to how well our findings would be represented at other institutions. Furthermore, due to the low number of interviewees, our findings are not necessarily representative of the broader graduate student community, here at the University of Colorado Boulder.

%%%%%%%%%%%%%%%
\subsection{\label{sec:interview}Interviews}
The interview protocol consisted of three different sections: one focusing on the context of the participant's research lab, another on the details of their documentation, and a third on their educational experiences in their undergraduate and graduate programs.

Section 1 contained questions addressing the broad context of the lab in which the interviewee did their graduate research. This section was intended to verify that the interviewee was in a research environment that necessitated the type of record keeping that we were interested in probing (e.g. a student who expresses that although they are in an experimental research group, their work consists solely of computer-based numerical simulations, would not be included in the study). 

Section 2 of the interview focused on the details of the interviewee's documentation in their lab notebook. In addition to answering questions, the interviewee would go through their notebook with the interviewer in order to provide examples to support the interviewee's answers. From these responses, we could determine if the interviewee's scientific documentation was consistent with our broad expectations about what constitutes lab notebook use (i.e., their records displayed the characteristics we outlined in Section \ref{sec:introduction}). 

Section 3 focused on the interviewee's training and experiences in using a lab notebook, over the course of their physics education. The purpose of this section was to obtain detailed descriptions of the experiences with scientific documentation that the interviewees had from their undergraduate education up to the present. We addressed both formal and informal experiences as well as those that occurred in a course and/or research environment. The interview protocol was the result of several iterations and discussions between both authors. This section comprised the majority of data source used for our analysis. The structured portion of Section 3 is as follows.
\emph{\begin{enumerate}
\item{``Did you, at any point in your education, receive instruction/training in how to keep a lab notebook? If so, can you elaborate on the context of this training?''}
\item{``Do you think this training sufficiently prepared you to maintain a lab notebook for your research in graduate school?''}
\item{``Do you think there was anything missing from this training that you should have learned?''}
\item{``When joining a research group did you feel you had to change/adapt the way in which you kept a lab notebook? If so, how?''}
\item{``Did you receive any instruction in maintaining a notebook from your research group, in graduate school?}
\item{``What feedback, if any, do you receive from your [graduate advisor] about the way in which you keep lab records in your lab notebook?''}
\item{``At what point do you recall coming to understand the importance of maintaining a lab notebook for scientific research?''}
\item{``What is your evaluation of the quality of your own notebook keeping practices? Do you feel it is suitable for your research?''}
\end{enumerate}}

 Questions 1--4 were intended to help answer our first two research questions, outlined in Section \ref{sec:introduction}, given that they probed the experiences students had during their undergraduate education and asked them to evaluate these experiences. Questions 5 and 6 were intended to answer our third research question, given that they directly addressed the experiences students had with the members of their graduate research group. Finally, questions 7 and 8 helped to address our fourth research question, by asking the interviewees to directly evaluate their current practice.
 
All interviews, which were conducted by author one, lasted between 50--70 minutes, and were semi-structured: the above listed questions (along with the remaining portion of the interview protocol) were asked of all interviewees. Then, clarification and follow-up questions were asked depending on flow of the interview. As an example, when asked \emph{``Did you, at any point in your education receive instruction or training...''} if the interviewee only described experiences in an undergraduate lab class without explicating the absence of training during, say, an undergraduate research experience then the interviewer would follow-up by asking \emph{``Were there any other experiences you had with lab notebooks---for example, an undergraduate research experience?''} In this way, we could explicitly establish both the experiences the interviewees did and did not have in the use of lab notebooks.

%%%%%%%%%%%%%
\subsection{\label{sec:pre-code}Preliminary Analysis}
For the purpose of this study, Sections 1 and 2 were evaluated qualitatively and in aggregate---they served to evaluate whether or not an interviewee's responses were appropriate for inclusion in the pool of subjects to analyze in detail.  Although these two sections are rich data sets, detailed analysis of them was not within the scope of this paper (we are currently writing a follow-up paper on these other two sections that outline features of authentic scientific documentation, which could be used as guidelines for lab instructors to include authentic notebook use in their lab activities). Of the 16 graduate students interviewed, 2 were removed because their undergraduate education did not conform to a conventional physics bachelors program and 1 was removed because their graduate work could not be classified as one of the ``table-top'' experimental sub-disciplines described above. 

The remaining 13 interviews were then coded in detail, as described in the subsequent section. Section 3 was the only section that was systematically coded and analyzed. Any subsequent reference to the interview data is in reference to Section 3

%%%%%%%%%%%%
\subsection{\label{sec:coding}Interview Coding and Analysis}
The goal of the coding was to capture the range of different experiences the interviewees had with using lab notebooks throughout the stages of their education. For this reason, we were interested only in the presence or absence of the various experiences as well as in what environment they did or did not occur (\emph{i.e.}, whether the code was or was not present at a particular point in their education). The multi-stage coding process incorporated both \emph{a priori} and emergent coding. The coding was done in three phases using the NVivo software. 

All detailed coding was performed by author one. In order to corroborate the coding, both authors regularly discussed the evolution of the coding scheme and examined specific examples of the codes to establish that they were reflective of the data. Given that the codes were simple to interpret and the conclusions we reached were not complex or subtle, we determined that it would be sufficient for one person to perform the coding.

The goal of the first phase of was to outline the different stages of the interviewee's narrative---it consisted of coding (1) the questions asked by the interviewer (to delineate the different topics) and (2) the different stages of education, discussed by the interviewee. This was done to segment the interview for easier analysis. This first phase was \emph{a priori}: we assumed the interviewee would be referencing only three different educational stages---namely, (1) undergraduate lab course experiences, (2) undergraduate experiences that weren't courses (\emph{e.g.}, undergraduate research experiences (URE), internships, alternate employment), and (3) graduate research experiences (including both Masters and Ph.D. graduate programs). In some circumstances, the interviewee would compare and contrast their undergraduate and graduate experiences---these instances would be coded with both stages of their education. This initial coding allowed us to isolate the different periods in the interviewee's education that facilitated the coding of the specific experiences, which was done in the subsequent phase. 

It should be noted that we chose to focus only on non-introductory lab courses in our analysis. A number of the interviewees discussed their introductory labs, but we chose not to focus on these. This was done to narrow the scope of the time period we were addressing. Furthermore, we assumed the non-introductory lab course experiences were more germane to the type of documentation on which we are focusing.

The second phase consisted of emergent coding of the various experiences described by the interviewees as well as their evaluation of these experiences. This process was emergent in that new codes were created as novel experiences and evaluations arose during the course of listening to each interview. Prior to coding each interview, we listened to the audio and wrote a brief narrative of the interviewee's educational arc, starting from the interviewee's description of their earliest experiences with using lab notebooks. This narrative acted as a snapshot of the individual's story and aided in the subsequent coding process. This narrative helped to approximately determine if the interview contained any novel experiences for which codes had not yet been established or if the narrative was consistent with previously coded interviews. We coded for both experiences (\emph{e.g.}, receiving formal or informal training, grading of notebooks in lab classes, receiving feedback from supervisors) as well as interviewees' evaluations of these experiences (\emph{e.g.}, feeling more feedback would be helpful; feeling that use of notebooks in their lab class was/was not similar to their use in research; use of lab notebook sufficiently prepared them for record keeping in research). Lastly, a \emph{Quotations} code was used to label interesting and well-phrased statements. What constituted ``interesting'' was subjectively determined while listening to the interviews. Uses of the \emph{Quotations} code were almost universally coincident with one or more of the experience and evaluation codes. The use of this code was not motivated by a particular type of experience the interviewees described---rather, it was used to capture the range of different experiences in the interviewees' own voice. These statements were transcribed and many of them are presented in Section \ref{sec:results} as specific examples of the broader results.

Once all interviews had been coded in this manner, the different experiences and evaluations were combined and consolidated into a smaller set of codes, each of which encompassed a broader and more general segment of the landscape of experience described by the interviewees. This process was done by identifying experiences or evaluations that were thematically similar to one another and grouping these together. These consolidated groups were then given a definition that best encompassed the range of individual descriptions (see Table \ref{tab:codes}). For example, one of the final ``Experience'' codes was ``Instruction,'' which was the consolidation of the following: ``verbal training,'' ``written guidelines,'' ``explicit guidelines,'' ``vague guidelines.'' An example of one of the final ``Evaluations'' codes was ``Beneficial,'' which was consolidated from: ``helped individual understand importance of notebook,'' ``improved notebook practice,'' ``authentic notebook practice,'' ``practically useful for lab activity.''   These groups then became the final set of codes. In the final phase of coding the consolidated coding scheme was applied to all of the interviews.

This final set of codes, along with the definitions, are listed in Table \ref{tab:codes}. The types of experiences that the interviewees described broke down into three broad categories: (1) \emph{instruction}---the individual received some kind of instruction or training in how best to maintain their notebooks from an external source, (2) \emph{practice}---the individual underwent self-guided use of a lab notebook for both recording and referencing lab records, and (3) \emph{feedback}---the individual received some form of written or verbal feedback on their practice with using a lab notebook for their lab records (to be classified as \emph{feedback} the individual providing the feedback must have actively looked through the notebook). The interviewees' evaluations of these experiences could be categorized in three ways: (1) it was described as being essential to developing documentation skills that were adequate for research; (2) they spoke positively about it, stating that it was beneficial to improving documentation skills; or (3) they spoke negatively about it, stating they did not feel it was beneficial to improving documentation skills.
%% CODE DEFINITIONS
\codetable

%%%%%%%%%%%%%%%%%%%%%%%%%%%%%%%%%%%%%%%
\section{\label{sec:results}Results and Discussion}
Each interviewee described an educational progression from undergraduate courses to graduate research that was unique, in many respects. Although no two educational paths were exactly the same, there emerged many commonalities in their experiences using lab notebooks. In addition to the results of the final phase of coding (which can be seen in Table \ref{tab:coderes} with the code definitions in Table \ref{tab:codes}), we discuss several of the broad themes that were observed in the interviews, as well as more specific descriptions of how these different themes manifest. Specific examples are highlighted by interview excerpts. In order to maintain the privacy of the individuals, some of the excerpts have been edited to remove any identifiable information. Additionally, in order to improve understandability of some excerpts, words have been changed or added. Both of these edits are denoted by square brackets. All edits are consistent with the Modern Language Association prescription for the treatment of quotations, and were done in a manner that preserves the meaning of the excerpts. 

%%%%%%%%%%%%%%%
\subsection{Undergraduate Lab Courses} 

One of the major themes expressed in the interviews centered on the interviewees' experiences in non-introductory undergraduate lab courses: these courses did not contribute much to their learning scientific documentation skills. Specifically, 11 of the 13 interviewees (I03 through I13, seen in Table \ref{tab:coderes}) described that their experience with notebooks did not adequately prepare them to use a notebook for lab records in research. They received no useful instruction or feedback on their use of a notebook, and only a few of them actually found their self-guided attempts at documentation beneficial. Furthermore, eight of the interviewees (I06 through I13) either did not maintain a notebook at all in these courses or, for those that did have a notebook, found the experience not to be beneficial for documenting their lab activities. Only one (I01) described having a comprehensive experience (that included instruction, practice, and feedback) in learning how to use a notebook during these courses. Furthermore, they described these as being influential in helping develop their current scientific documentation practices.

Interviewees articulated a range of ways in which their lab courses did not address how to use a notebook in a productive way. One common criticism (made by seven of the interviewees) was that the design and implementation of the lab curriculum and lab activities themselves did not actually necessitate the use of a lab notebook. This was explained as largely being due to the short timescale of their lab activities, as described in the following quotes by two of the interviewees.
\begin{itemize}
\item[]{\emph{``In undergrad labs there is not a real continuity from day to day. You do one experiment and finish, so there seems like less of a pressing need to record everything, because you do it all that day.''}}
\item[]{\emph{``I remember from the lab [classes], thinking that I'm never going to need to remember any of this after the next week. Because you go in for one week and do an experiment and then you never [revisit it]. That experiment was done and you write your lab report.''}}
\end{itemize}

Similarly, seven of the interviewees who described using a notebook in their lab class, expressed that this was not reflective of how notebooks are used to keep lab records in a research setting, and therefore did not provide a particularly useful learning experience. This view can be seen in the following quotes.
\begin{itemize}
\item[]{\emph{``All of the reasons I use a lab notebook now did not apply to the experiments I did in that class. The reason is because the experiments were scripted and they were meant to be contained in an afternoon."}}
\item[]{\emph{``[T]hey had us keep lab notebooks ... but it was very different from how I would actually keep a lab notebook. Essentially, since those lab activities were basically just worksheets they would tell you what to do and you'd do it and write about it. I would say that wasn't very relevant [to learning how to keep a notebook] for research.''}} 
\end{itemize}

Another criticism of the lab courses that was echoed by three interviewees was that the instruction, feedback, or grading focused on superficial and inauthentic aspects of the notebook. This is made clear in the following examples, quoted from two of the interviews.
\begin{itemize}
\item[]{\emph{``[The lab course] was the opposite of an authentic experience ... we were preoccupied by writing all of the check-box things that they would grade on … the notebook served no other purpose than to jump through the hoops that were required to get a good grade."}}
\item[]{\emph{``There was a little too much emphasis on there's these elements that you need to check off and you'll be good, and not enough emphasis on thinking about what you should be doing."}}
\end{itemize}

On the other hand, four interviewees stated that there was no requirement to use a lab notebook in their classes nor emphasis on their importance for doing laboratory science. As a result, many students did not keep one. 
\begin{itemize}
\item[]{\emph{``I was ready to come in and follow directions ... Yeah, that was my experience, and I didn't even keep a lab notebook. The only thing we had to do was turn in four lab reports."}}
\item[]{\emph{``We had a multi-page guide for the lab [we] were working on. A lot of [the records] would end up in the margins. I didn't have a notebook for the class. I don't recall writing in something very conscientiously.''}}
\end{itemize}

Additionally, the lab course instructors did not provide any guidelines or feedback as to how scientific documentation could be performed. In most cases, the individual's grade was mostly based on summative lab reports and not on any of their daily notebook records.

As a general answer to our first research question (What experiences did they have with lab notebooks in their non-introductory undergraduate lab courses and how do they evaluate the quality of these experiences?), this section suggests the majority of the interviewees did not develop documentation skills as a result of the time spent in their lab courses. The interviewees did not feel that the structure of the lab courses were conducive toward learning this skill. However, it is unclear whether or not the development of these skills was actually an explicit learning goal of the courses. In contrast, the two interviewees (I01 and I02) who did describe a lot of benefit from their lab courses, were two of three interviewees to evaluated their current documentation methods as ``high quality'' and ``thorough.'' 
%% CODING RESULTS
\resultstable

%%%%%%%%%%%%%%
\subsection{Graduate Research}
Another commonly articulated theme centered on the interviewees' experiences in their graduate research group: all 13 interviewees stated that they received little to no instruction or training from their graduate advisors in how they should maintain these lab records for their research. Furthermore, 12 (all but I02) stated that their advisors rarely, if ever, provide any feedback on the quality of their notebook or would rarely look through the graduate students' lab notebooks, if at all.

The extent of the instruction or feedback, if any, consisted of the advisor making an occasional suggestion about a particular parameter that should be written down or suggesting that the students should ``write more.'' In general, 10 interviewees describe that their advisors were unaware of the state of the lab notebook records, given that they rarely looked at them.
\begin{itemize}
\item[]{\emph{"Not really, [my advisor] was pretty hands off, in a way. Any relevant information [my advisor] needed from me they would ask me for it, but not necessarily look in my lab notebook."}}
\item[]{\emph{``The original instructions I got were fairly minimal and didn't go much beyond writing down the date and what I do.''}}
\item[]{\emph{``My [advisor] never sees my experimental or personal [notebook], so [my advisor] gives no feedback or input on this.''}}
\item[]{\emph{``Nope, [I got no feedback]. They generally don't see my notebooks. When I present data to them, I make a power point presentation with the relevant information.''}}
\end{itemize}

In light of this lack of training, three stated that they felt ill-prepared to keep effective lab records when starting graduate research---they did not know what level of detail they should write, what things were important to write down, nor how to structure the information, as seen in the following quotes.
\begin{itemize}
\item[]{\emph{``I didn't want to do it. Part of it was difficult because I didn't know, what was important to write down. So it just seemed so overwhelming. What am I supposed to do, write down everything?''}}
\item[]{\emph{``I got this notebook on the first day and [my advisor] said 'Write important things in it.' I was very nervous, because I had no idea what I was doing ... I [was] freaking out about this pressure and I think that's what I see in the beginning [of my notebook]."}}
\end{itemize} 

However, four interviewees indicated that they had received some informal training from group members, other than their advisors, that was beneficial to the development of their scientific documentation. These group members were either post-docs or senior graduate students. These experiences were described as being relatively collaborative, in that the interviewee also had an active role in establishing documentation norms.

Broadly speaking, this section provides an answer to our third research question (What kinds of guidance, practice, or feedback did they receive from their advisors or colleagues in their graduate program?): the interviewees' research advisors played a minimal role in developing the interviewees' documentation skills. With that said, it is unclear what the advisors' perspectives are of the graduate students' ability to perform scientific documentation.

%%%%%%%%%%%%%%%%
\subsection{Developing scientific documentation expertise}
Despite the lack of oversight and structure that they experienced with lab courses and research advisors, nine (I01, I02, I05--I07, I09--I11, I13) felt that their current documentation practice was sufficient for their research---answering our fourth research question (How do they evaluate the current state of the records they document?). Given that they felt they had eventually developed these skills, one may then ask, how and when did they learn them?

In broad terms, interviewees described learning how to use a lab notebook mostly through an informal self-guided process involving a substantial amount of hands-on practice in documenting authentic scientific records. This practice occurred in one of three environments: (1) an undergraduate research experience (URE), (2) a stand-alone Master's program, or (3) in their current graduate research group. Of the 13 interviewees three described having very effective undergraduate research experiences where they were able to gain a lot of practice with scientific documentation. This resulted in them feeling prepared to maintain effective records in graduate school. Additionally, two of the interviewees had separate Masters programs prior to their current doctorate research in which they felt they had sufficient practice to prepare them for their current research. Lastly, five interviewees only began to develop these skills effectively once they had already started in their current Ph.D. program. Six interviewees felt that their current practice still requires improvement.

Seven of the interviewees expressed that they recognized early on (sometime during their undergraduate education) that documentation skills were valued in the scientific process, but actually understanding the specifics of what good documentation entailed and how to put that into practice did not come until later, when they were able to utilize it in a research setting---conveyed in the following, by three of them:
\begin{itemize}
\item[]{\emph{``I feel like I understood fairly early on in undergrad (maybe after my first or second lab class) that lab notebooks were very important. But I don't feel like I understood how to keep a good lab notebook until after my first year of grad school.''}}
\item[]{\emph{``I theoretically understood from the first time I was told to keep one in an advanced lab class but wasn't committed to it yet...[it was] during my undergraduate [research when I] fully realized the importance. When I had to retrace my steps in order to find out enough about my experiment to actually do a comprehensive analysis of my data.''}}
\item[]{\emph{``I could always see why you would want to keep good records, but I think that it's important to do it, to understand it fully. I didn't really appreciate what `good records' meant until I kept them.''}}
\end{itemize}

All interviewees emphasized the importance of hands-on experience and authentic practice maintaining a lab notebook (voiced in the previous quotes) in order to develop effective documentation skills. As seen in Table \ref{tab:coderes}, nine of the interviewees had hands-on practice with using notebooks that helped improve their abilities, in an undergraduate research setting. Note that all but I12 and I13 lacked any kind of instruction or feedback from their advisors in this setting. All interviewees stated having hands-on practice during their graduate research. 

These hands-on experiences came in a range of different forms, but the predominant experience consisted of extended periods of trial and error. This sentiment was universally described.
\begin{itemize}
\item[]{\emph{``[I]t took me a year or more to work out for myself what a lab notebook was ... it was an art to realizing of all the things going on that one could write down, what you really needed to write down. I don't think there's any way I could have learned that, beyond just the experience of having going through it for a year or so.''}}
\item[]{\emph{``You start learning what's useful and what might not be useful, but there's always something you have no idea will be helpful. It's hard to learn what those are until you've been working with it for a while.''}}
\item[]{\emph{``Now I have a pretty good idea of what is important and what's not, [but that developed] very slowly over time. From that experience I had some episodes in the lab where something would work and I'd understand it, so then I would write it up very carefully ... After some time had passed I had reason to revisit the records, I was very glad I'd kept them.''}}
\end{itemize}

Three had these hands-on experiences through reading old lab notebooks that had been maintained by previous researchers in their group. By attempting to learn about the previous progress on their experiment they recognized what pieces of information were crucial to making further progress and ways to effectively convey that information. This was then incorporated into their own lab notebook documentation, as made clear in the following.
\begin{itemize}
\item[]{\emph{``As I started to look back on old lab notebooks being like `I want to know if they did this thing,' then I started realizing all the things I'd want to look back on in [my notebook].''}}
\item[]{\emph{``I also had my first experience with referring to the lab notebooks of former students to find relevant information that might be useful in my work; the excellent lab notebook examples provided there were an unofficial model for me in my own notebook.''}}
\end{itemize}

Three interviewees had negative experiences that led to their understanding of how these records should be kept---by not writing down various pieces of important information, they prevented themselves from making further progress in their projects. For these individuals, these experiences had an impact on how they chose to maintain their notebooks, going forward. The following quotes describe just such experiences.
\begin{itemize}
\item[]{\emph{``[During my undergraduate research experience,] I was taking data by myself. I forgot to write down [some parameters], so I had things written in all kinds of wrong spots. It was some of the first few hours of completely wasted time in the lab that I experienced.''}}
\item[]{\emph{``Most of my procedures I have adapted myself from experience, mostly negative experiences where I needed information than I hadn't written down.''}}
\item[]{\emph{``[I]t's more clear to me what kinds of things need to be well organized, so that I don't screw myself over six months from now ... there were moments where I was like `shoot, I should have wrote that down.' But I think there's less of those as I gain more experience.''}}
\end{itemize}

Three of the interviewees described that there was a general lack of scientific documentation infrastructure in the labs they joined when starting research in graduate school and therefore had to be proactive in developing documentation procedures for themselves, as well as for others in the group.
\begin{itemize}
\item[]{\emph{``[When joining my research group], my senior grad student didn't write many things down and that caused a lot of headache. I think I helped pushed us to be a little bit more conscientious about writing stuff down and a lot of it is me doing the writing.''}}
\item[]{\emph{``When I came [to grad school], in the lab there was very little record keeping. One person did keep records and kept it in their own personal notebook. Then we started developing the [collaborative] experimental ones and I tried to start writing in my own more."}}
\end{itemize}

The major emphasis by the interviewees was that having hands-on practice is the central way to develop good documentation skills, and this practice occurs primarily in authentic research settings. This is made clear by Table \ref{tab:coderes} which highlights that the majority of useful practice occurred in the undergraduate and graduate research settings. However, this practice was not typically accompanied by instruction or feedback from their supervisors. 

Broadly speaking, this section answers our second and fourth research questions (Did they have any other experiences with scientific documentation during their undergraduate education? How do they evaluate the current state of the records they document?): Many of the interviewees took part in undergraduate research experiences in which they got practice with documentation, but most did not have any guidance or instruction in this setting. By the time of the interview, most of the interviewees felt their current records were sufficient for their research, however a number of them felt they should improve their documentation in some manner.

%%%%%%%%%%%%%%%%%%%%%%%%%%%%%%%%%%%%%%
\section{\label{sec:disc}Summary and future directions}
Through interviews of experimental physics graduate students, we explored how these researchers developed a key scientific skill---scientific documentation. Though the landscape of experiences is varied, several common themes emerged from the interviews that addressed our research questions. These themes are summarized in the following three paragraphs.

The first theme was that most of the interviewees indicated their upper-division undergraduate lab courses were not a major contributor in learning how to perform scientific documentation. Furthermore, many found that documentation wasn't necessary or useful for the types of lab activities they were doing in these courses, and thus didn't keep a notebook of any kind.

The second major theme was that the interviewees generally lacked guidance in scientific documentation, from their research advisors during graduate school. Even when starting in the lab, the interviewees described getting essentially no instruction from their advisors. Any external input the interviewees received came informally from their colleagues (senior graduate students or post-docs) and was typically minimal. The research advisors rarely, if ever, directly inspected the information that the interviewees recorded in the lab notebooks. A number of the interviewees felt ill-prepared to keep such records at the outset of their graduate research and described their early notebook keeping as lacking.

However, the third theme we found was that most of these graduate students did eventually develop scientific documentation skills that they felt were adequate to succeed in their research. The predominant factor in this development was a great deal of time spent with informal hands-on practice in an authentic research setting. The interviewees described getting these hands-on experience either during undergraduate research experiences or during the first couple years of their graduate research.

These findings suggest some questions that would require further inquiry: Firstly, are documentation skills actually explicit learning goals of undergraduate lab courses and if so how are these courses attempting to teach these skills? If indeed this is recognized as an important learning goal, then how should a lab course be designed to most effectively teach documentation? Additionally, what are the expectations of research advisors for the scientific documentation in their research labs, and how aware are they are about the state of the documentation being recorded by their graduate students?

To address the first two question, interviews with lab instructors could be carried out to establish how much lab notebook use is believed to be an important part of these courses. As suggested by a number of the interviewees, notebooks need to be made an essential and authentic part of the lab activities themselves, in order to have a more effective use of documentation skills in these courses. We speculate that this could be done in a number of ways. For example, including multi-week projects as a part of the course curriculum would make it necessary for students to rely on their documentation to keep track of progress. Additionally, lab activities could be designed to necessitate collaboration between multiple students in such a way that each student must rely on the documentation their peers record in their own notebooks or in a collaborative group notebook. In both of these examples, the lab context would change in such a way to more accurately reflect an authentic research setting. These changes would require thoughtful framing on the part of the instructor, and further research needs to be done in order to determine how best to approach such a course transformation. 

In order to address the second question, interviews with research advisors could be performed to shed light on their beliefs about their graduate students' abilities with scientific documentation, as well as how they might best support their graduate students to further improving their existing practice. These interviews could even help inform how to incorporate notebook use in lab courses that would be most in-line with the expectations of graduate research.

In conclusion, it is important to note that most of the interviewees had developed documentation methods they felt were sufficient for their research. Generally, they did this without the external support of instruction and feedback. But, it is reasonable to believe that by providing instruction and feedback in the undergraduate lab course setting, as well as a graduate research setting, the interviewees may have developed their documentation skills sooner and would more readily adapt to a research environment. 

%%%%%%%%%%%%%%%%%%%%%%%%%%%%%%%%%%%%%%%%
\begin{acknowledgments}
We would like to acknowledge and thank all of the graduate students who took part in this study as well as Dimitri Dounas-Frazer and Bethany Wilcox for their helpful feedback. This work was supported by NSF grant nos. DUE-132-3101 and DUE 133-4170.
\end{acknowledgments}

%%%%%%%%%%%%%%%%%%%%%%%%%%%%%%%%%%%%%%%%

%

%%%%%%%%%%%%%%%%%%%%%%%%%%%%%%%%%%%%%%%%
\end{document}